\documentclass[12pt]{article}
\usepackage{natbib}

\usepackage{amsmath}
\usepackage[margin=1.3in]{geometry}
\usepackage{hyperref}
\usepackage{graphicx} 

\title{Another 100 Years of Quantum Interpretation?\footnote{An edited version is to appear in: \textit{How to Understand Quantum Mechanics – 100 Years of Ongoing Interpretation}, edited by Lars-G\"{o}ran Johansson and Jan Faye.}}
\author{Karen Crowther\\\textit{Department of Philosophy, Classics, History of Art and Ideas},\\ \textit{University of Oslo}}

\begin{document}

\maketitle

\bibliographystyle{apalike}

\begin{abstract}\noindent

Interpretation is not the only way to explain a theory's success, form and features, and nor is it the only way to solve problems we see with a theory. This can also be done by giving a reductive explanation of the theory, by reference to a newer, more accurate, and/or more fundamental theory. We are seeking a theory of quantum gravity, a more fundamental theory than both quantum mechanics and general relativity, yet, while this theory is supposed to explain general relativity, it's not typically been thought to be necessary, or able, to explain quantum mechanics---a task instead assigned to interpretation. Here, I question why this is. I also present a new way of assessing the various interpretations of quantum mechanics, in terms of their heuristic and unificatory potential in helping us find a more fundamental theory.

\end{abstract}


\section{Introduction}

Interpretations of physical theories serve many purposes. For quantum mechanics, two of these purposes include \textit{solving problems}---particularly the measurement problem---we may see as associated with the theory, and \textit{providing explanation} of the theory. An interpretation should facilitate an understanding of how the theory works, why it works, and why it has the form and features that it does. \footnote{Often, interpretation is thought to do this by giving us insight into what the theory tells us about nature---what the world could be like, if the theory were true. Such an explanation allows us to picture ``what's really going on'' according to the theory, which many philosophers argue is necessary particularly in quantum mechanics. But this ``picturing of nature'' is not a requirement on an interpretation. As Healey emphasises, an interpretation need not assume that the theory offers representations and/or descriptions of the physical world. An interpretation can explain how and why the theory is successful without requiring a ``realist'' (i.e., \textit{representationalist}) perspective \citep[][p. 730]{Healey2012}. Thus, while, traditionally, interpretations of quantum theory have been broadly classified as either \textit{realist} or \textit{non-realist}, there has been a shift towards now distinguishing between \textit{representationalist} and \textit{non-representationalist} interpretations, see \citet{Healey2020} and \citet[][fn. 2, p. 79; 87]{Wallace2020}. Non-representationalist interpretations can and do still give us understanding of the theory, as well as allow us to utilise the theory and to make sense of its results. So, whether representationalist or not, interpretation is tied to explanation of the theory, and interpretation can be invoked in order to solve problems we see with the theory. To be clear, ``shut up and calculate'' forms of anti-realism are not interpretations on this account.}

Interpretation, however, is not the only means of explaining how a theory works, why a theory works, and why a theory has the form and features that it does. And nor is interpretation the only way of solving problems we see with the theory. A different way of achieving explanation of a theory---rather than by interpretation---is by appealing to inter-theory relations, providing a \textit{reductive explanation}.\footnote{Reductive explanation is distinct from interpretation, but is connected to it. Our reductive explanation can be influenced by our interpretation of both the reducing and reduced theories (e.g., our interpretation tells us what needs to be explained and what does not, as well as how to understand the link between ontologies of the two theories). And, also our interpretation of the theory can (and standardly will) be affected by the knowledge we gain from the reductive explanation of it (the most obvious example of this being when the reductive explanation forces us to re-interpret our theory as an \textit{effective theory}, valid only in a limited domain).} A theory may be successful and have the features that it does because it is an approximation to another theory which is more fundamental, or otherwise more accurate in describing the world (or, for non-realists, more accurate in making predictions and allowing us to successfully interact with the world). Here---carefully---I mean to distinguish between these two different possibilities: in the first case, where our theory approximates a more fundamental theory, we can have an inter-level reductive explanation, while in the latter case, where our theory approximates a more accurate theory formulated ``at the same level'' of fundamentality, we can have an intra-level reductive explanation. Here, I refer to the inter-level relationship as \textit{vertical reduction}, and the intra-level relationship \textit{horizontal reduction} or \textit{successive reduction}. The ``levels'' here refer to levels of fundamentality. I explain what I mean by ``more-fundamental theory'' below (\S\ref{sect:fund}), but for now, it's that the theory is more universal, more unificatory, and has a broader domain of applicability than another (whose domain is fully, or mostly, contained within the domain of the more-fundamental theory).

Next, in regards to solving problems we see with a theory, another way of doing this---rather than by interpretation---is by appeal to a newer, more accurate theory. In the case of general relativity, for instance, we typically hope to solve its problems of, e.g., indeterminism associated with spacetime singularities, by moving to a theory of quantum gravity. In the case of quantum mechanics, we may hope to solve the measurement problem by moving to a new \textit{approach} to quantum mechanics, such as a hidden-variables theory, or dynamical-collapse theory, which, by supplementing or modifying standard quantum mechanics may give us a newer, more accurate theory (which may be a more descriptively accurate theory rather than a more empirically accurate one---i.e, that it better captures the quantum ontology, but makes no predictions that differ from standard quantum theory) that solves the problems we had with the standard theory. In the first example, the problems are solved by a newer, more accurate theory which is also a more fundamental theory than general relativity. And, the explanation of the success, form and features of general relativity is provided via vertical reduction (as well as horizontal reduction). In the second example, the problem is solved by a newer, more accurate theory which is \textit{not more fundamental}, and the success, form and features of the older theory (standard quantum mechanics) being explained via (only) a horizontal reduction.

Note that I refer to the hidden-variables theories and dynamical collapse theories as ``approaches'' to quantum mechanics rather than ``interpretations''. This is because, strictly speaking, \textit{interpretation} should not modify or provide additional theoretical structure to the standard theory. (Most of) the hidden-variables theories and dynamical-collapse theories do either supplement or modify standard quantum mechanics, and are thus better thought of as \textit{approaches} or \textit{alternatives} to quantum theory. Here, though it's important to emphasise, following \citet{Wallace2020}, that quantum theory is really a \textit{framework} theory, and most dynamical-collapse and hidden-variables approaches do not engage with the framework as it is actually applied across different levels, but rather ``supplement or modify one \textit{specific} quantum theory: nonrelativistic quantum particle mechanics'' (p. 93).

If these approaches (whose dynamical equations differ from those of standard quantum theory) are successful in their aims of providing a conceptually and physically acceptable account of quantum phenomena, in recovering non-relativistic quantum particle mechanics, and are shown to be empirically successful, they can potentially give us reductive explanation of non-relativistic particle mechanics. As stated, this would be intra-level reduction, since the newer, more accurate theory is not more fundamental than non-relativistic quantum particle mechanics, being neither more universal nor unificatory---e.g., it is still not a relativistic theory. Further, these approaches, even if they were successful in giving us intra-level reductive explanation of \textit{the framework} of quantum (field) theory (rather than simply non-relativistic quantum particle mechanics), would not offer us explanation of quantum theory in terms of a more fundamental theory. \textit{Why is it that, in solving the measurement problem and seeking an explanation of quantum theory, we overwhelmingly look to interpretation or alternative approaches rather than seeking a deeper explanation in a more fundamental theory?} This question is particularly striking because, 1) we are looking for a more fundamental theory anyway, and 2), we standardly think that the problems with general relativity, for instance, are to be solved by this more fundamental theory. In addressing this question, I develop a new account of relative fundamentality of physical theories that helps illuminate the problem more sharply.

There is utility and purpose to interpretations, and also for alternative approaches to quantum theory. In this paper, though, I suggest that the alternative approaches that seek to connect with a more fundamental theory offer \textit{additional utility and potentially deeper explanatory power} compared to approaches which do not. Interpretations, on the other hand, \textit{do not offer us this type of additional utility nor the potential for deeper explanatory power}, no matter whether they connect to a more fundamental theory or not. By this, I mean \textit{not} just those approaches which connect with a more fundamental theory than non-relativistic quantum particle mechanics, but---more radically---a more fundamental theory than either the Standard Model of QFT or,---especially---the framework of quantum (field) theory itself. This is because we have reason to believe that the framework of quantum theory, like our specific quantum theories, including the Standard Model, is not fundamental. Yet, searches for completions of the Standard Model and searches for quantum gravity have overwhelmingly assumed that the more fundamental theory will be a quantum theory, rather than a theory that reductively explains quantum theory. 

Why is this? I don't mean it as a sociological question; I'm interested in the physical and philosophical arguments. But, for now I can only speculate about these. What we can see is that interpretations, such as the many-worlds interpretation, are committed to taking quantum theory as given, and---apparently also---as given fundamentally. Interpretations offer a shallower explanation of quantum theory than approaches which seek to explain quantum theory by a more fundamental theory. They also offer only limited heuristic potential in helping us find a theory of quantum gravity, while restricting the form that such a theory could take. If the framework of quantum theory is taken as fundamental, then quantum gravity must fit within it. I do not mean to dismiss interpretations as worthless. \textit{My aim here is rather to give some positive argument for why we might be interested in being more radical than just interpreting quantum theory}. We could spend another 100 years debating quantum interpretations. Or, we could (perhaps in addition to spending another 100 years debating interpretations) spend some more resources in looking ahead, to a theory of quantum gravity that \textit{explains} quantum theory in a deeper sense than mere interpretation is capable of.

To begin, \S\ref{sect:fund} presents the applicable notion of relative fundamentality in physics, which allows us to distinguish between more-fundamental and less-fundamental theories. Along with this, the two ideas of reduction mentioned above are elaborated. The two sections following this explore why it is that our best and most-successful theories of physics---general relativity (GR) \S\ref{GR} and the Standard Model of quantum field theory (QFT) \S\ref{QFT}---are not thought to be fundamental. The more fundamental theory is quantum gravity, introduced in \S\ref{QG}. The measurement problem, which standardly motivates quantum interpretations and various approaches as solutions, is introduced in \S\ref{measur}. In this section I evaluate the Bohmian approaches \S\ref{Bohm}, dynamical collapse approaches \S\ref{GRW} and the many-worlds interpretation \S\ref{Everett} in regards to their heuristic utility and potential for deeper explanation of quantum theory. Some reasons to seek a deeper explanation of quantum theory are presented in \S\ref{deeper}, and some reasons why we might be sceptical that there is any deeper explanation for quantum theory are swiftly rejected in \S\ref{sceptic}.

\section{Fundamentality and reductive explanation}\label{sect:fund}

My aim in this paper is to question the fundamentality of quantum theory, and to provide some arguments for why we should be interested in seeking an explanation of quantum theory by reference to a more fundamental theory. What is meant by a ``more fundamental theory''? 

The definition of relative fundamentality that I take here is that a more fundamental theory is one which is broader than another, in the sense of being more universal, applying to a greater range of empirical phenomena (in the actual world), and being more unificatory than another---and where the domain of applicability of the less fundamental theory is contained within (or at least is mostly subsumed by) the domain of applicability of the more universal theory. This understanding of relative fundamentality makes no reference to scale, and importantly does not assume or imply that the more fundamental theory is the more ``micro'' theory, nor that the the less fundamental is a ``macro'' theory.

Since the development of quantum mechanics, there has been a tendency to think of more-fundamental physical theories as being those that describe ``small stuff''. This idea was further entrenched by the development of QFT and the Standard Model, which led to ``fundamental physics'' being identified with \textit{high-energy physics}. It cemented the orthodox view in metaphysics, in which the micro-constituents of nature are more fundamental than the macro objects, properties, and phenomena whose existence they are (somehow) thought to be responsible for. Metaphysics remains fixed in this perspective: that there is some hierarchy of being, where the less-fundamental, derivative entities depend on more-fundamental ones, and the more fundamental entities are the more micro ones. We have an intuitive view, tied to this metaphysics, of the basic ``building blocks'' of nature, which somehow compose or constitute the larger objects and structures of the universe. If the more fundamental stuff is described by high-energy physics, then it is reasoned that the high-energy physical theories which describe this more fundamental stuff are more fundamental than lower energy theories, which describe derivative entities and structures that depend, asymmetrically, on the more fundamental ontology. I think this is an unfortunate mistake, and one which has been at least partially responsible for the belief that the quantum is more fundamental than other physical theories.

The most vocal proponent of the fundamentality of high energy physics was Weinberg. Weinberg speculated that relative fundamentality had ``something to do with [...] greater generality'' of physical laws---for instance, Newton's laws compared to Kepler's laws \citep[][p. 435]{Weinberg1987}. This greater generality also means that Newton's laws \textit{explain} Kepler's laws. Weinberg saw that more general, more fundamental scientific theories were explanatory of less fundamental ones. Famously, he claimed, further, that these chains of explanations from less fundamental theories to more fundamental theories traced a path towards higher energy theories---that the ``arrows of explanation'' seem to converge on the ``very small'' \citep[][p. 435]{Weinberg1987}. I think Weinberg was correct that more universal theories are more explanatory and more fundamental, but he was incorrect to assert that these explanations always lead us towards higher energy scales. In my view, GR may be regarded as equally fundamental as the quantum: it is equally general and explanatory, and GR has not been explained by quantum theory.

A theory can explain another via theory reduction. Here, I take a broad understanding of reduction as derivation. We can also understand reduction as construction, following \citeauthor{relicts} (Forthcoming, p. 5): Theory $T_t$ is reduced to $T_b$ if the equations, quantities, and variables of $T_t$ can be constructed from the equations, quantities, and variables of $T_b$. This broad understanding of reduction allows for the use of various inter-theory relations in establishing a reduction, including, e.g., approximations, idealisations, limiting procedures, and other correspondence relations such as those described in \citet{Crowther2020}. It holds between a pair of theories, $T_b$ and $T_t$, which are accepted as empirically successful and whose domains of applicability at least partially overlap. The reduction provides explanation of the success, form and features of $T_t$ by reference to this theory being an approximation to $T_b$.

There are two different types of theory reduction. This has been recognised by various authors, though the specific accounts of these two types of reduction differ quite significantly between authors. What is common in most of these accounts is the basic idea that \textit{intra-level} reduction holds between an older theory and its successor theory, while \textit{inter-level} reduction holds between a macro theory and a micro theory (i.e., theories applicable to different energy scales). Of course, a reduction may be both types at once, when we have a newer theory that is also a micro theory compared to the older theory which is reduced. A standard example of intra-level, or successional, reduction is non-relativistic mechanics reducing to special relativity in the limit of low velocities (note: the philosophers' convention here is in the opposite direction to the physicists', which says that special relativity reduces to non-relativistic mechanics in the limit). A standard example of inter-level reduction is thermodynamics reducing to statistical mechanics at a finer-grained level of description. Although this example is also a case of an older theory reducing to a newer theory, the point is to emphasise that this case is supposed to illustrate a ``higher-level phenomenological theory'' reducing to a ``lower-level, micro-theory'' which provides a mechanistic explanation of the higher-level phenomena in terms of the behaviour of micro-constituents (particles).\footnote{I mention these examples just as illustration: both have been much discussed and contested.} 

It is the inter-level type of reduction that has been the most discussed: this type of reduction is supposed to be more explanatory than intra-level reduction. \citet[][\S4]{Wimsatt2006}, for instance, refers to this as \textit{explanatory reduction}, seeing its aim as to provide a compositional, causal, or mechanistic explanation for some higher-level phenomenon (object, process, or behaviour) in terms of behaviours of lower-level micro-ontology. But, intra-level reductions are also explanatory: one of their key aims is to explain the success of the older theory by reference to the newer theory, where the older theory is thought to be strictly false, or less accurate, than the newer, more accurate theory which has been accepted as its replacement \citep[][p. 1449]{Crowther2020}. This hasn't been properly recognised; for instance, \citet{Nickles1973} denies that intra-level reduction (``reduction\textsubscript{2}'') ``involve[s] the theoretical explanation of one theory by another'', claiming ``[n]ot all reduction is explanation!'' \citep[][p. 185]{Nickles1973}. (Rather, Nickles sees the key role of intra-level reduction as being the \textit{justification} of the new theory. But, surely, in order to justify the new theory by reference to the old theory, we are appealing to the success of the older theory, and if the older theory is to be replaced by the new one, the explanation of this success---via the reduction---is key to the justification of the new theory). 

The lack of recognition of the explanatory roles of intra-level reduction, and the focus on reductive explanations as inter-level ones, providing micro-compositional, mechanistic, or causal explanations---I speculate---may have also played in to our thinking that quantum is more fundamental than GR (even in the absence of a reduction). There is also a widespread tendency to interpret Einstein's 1919 \textit{London Times} article \citet{Einstein1919}, as well, as saying that ``constructive theories''---such as the kinetic theory of gases, which build up a model of a complex phenomenon from a description of its more-basic constituents---as of superior explanatory power. Constructive theories are seen as complete. And, on the other hand, that ``principle theories''---such as relativity theory, which begin with a set of high-level empirical generalisations---as \textit{explanatorily impotent} \citep{Lange2014}. Principle theories, and especially relativity theory, are seen as incomplete \citep{Einstein1907}.

Yet, as \citet[][p. 452]{Lange2014} notes, principle theories consist of law-like generalisations that unify a host of facts, and unification under law-like generalisation has often been thought to be intimately connected to explanatory power. (\citet{Lange2014} further presents several remarks from Einstein which indicate that he did take relativity theory to be explanatory, as well as offering an alternative interpretation of Einstein's 1919-20 remarks which does not take principle theories to be explanatorily powerless).

It's time to move past our bias towards micro-constitutive explanations, and the view of theories describing micro-constitutive entities or processes as more fundamental than others. Accordingly, I propose a new distinction between inter-level and intra-level reduction, where ``levels'' refer to levels of fundamentality, and a theory is more-fundamental, or lower-level, than another if it is more universal and more unificatory. Here, again, $T_t$ and $T_b$ refer to theories which are empirically confirmed, and where the domain of $T_b$ either (mostly) overlaps with, or is subsumed by, the domain of $T_t$.
\\\\
\textbf{Vertical reduction} (inter-level): $T_t$ is a higher-level, less-fundamental theory that is reduced to a lower-level, more fundamental theory, $T_b$, which has a broader domain (applying to a greater range of empirical phenomena), is more universal and more unificatory. ($T_b$ = `bottom').\\\\
\textbf{Horizontal Reduction} (intra-level, or successional reduction): $T_t$ is an old, tainted theory that is reduced to a newer, better theory, $T_b$, which is more accurate. ($T_b=$ `better'). 

Both types of reduction are explanatory! This will be illustrated with the case of quantum theory---though, in this case we do not actually have the reduction in either sense, since the candidates for $T_t$ have not been fully developed or accepted as empirically successful, so I will be discussing only hypothetical reductions. Hypothetically, the success, form, and features of particular quantum theories can be explained by alternative approaches to quantum theory, such as hidden-variables approaches or dynamical collapse approaches, if these approaches succeed in deriving or constructing those particular quantum theories. And, arguably, if they are able to provide a standard recipe for doing this, consistent across all different specific quantum theories, they could also potentially explain the framework of quantum theory \citep[cf.][]{Wallace2020}. In either case, this would be a horizontal reduction, since these new theories do not have a broader domain (they describe the same range of empirical phenomena), greater universality, nor greater unification than the quantum theories they reduce, and thus are ``on the same level'' of fundamentality as quantum theory, in spite of their being more empirically and descriptively accurate than quantum theory. 

If we were to have a more fundamental theory of quantum gravity (\S\ref{QG}) from which the Standard Model of QFT were able to be derived, or from which we were able to construct the framework of quantum theory, then this would provide a vertical reduction that explains the success, form and features of either the Standard Model or the framework of quantum theory. Since quantum gravity would be a new theory, this reduction would also represent a horizontal reduction. \textit{I would like to make the case that this is a more useful and more explanatory goal to strive for than the work on approaches and interpretations of quantum theory that does not aim for a more fundamental explanation of the theory}. This is because the explanation goes deeper, it is unificatory and describes/applies to more of the world. Aiming for such an explanation also has heuristic utility in the search for quantum gravity, compared to approaches and interpretations at the same level as quantum theory, which are of limited utility in aiding the quest for the new theory.

But---importantly---there are different levels to this goal of providing a deeper, more-fundamental explanation. The first level is if the new theory of quantum gravity is itself a quantum theory, i.e., constructed via, or fitting with, the current framework of quantum theory. This has been the standard approach to quantum gravity, where (most of) the different approaches utilise different quantisation recipes to quantise gravity. In this case---if it were successful---we would get a deeper explanation of the success, form, and features of both GR and the Standard Model of QFT (by reduction to the more unified theory), but the `unsettling' aspects of quantum theory, including the measurement problem, would remain in need of explanation. Thus, in this case, there would still be an important role for quantum interpretation. An example is Rovelli's relational quantum mechanics---an interpretation applied to loop quantum gravity \cite{RQM, Healey2021}.

Alternatively, we could strive for deeper explanation still---one where quantum theory itself is reduced to a more fundamental theory. This would represent the most complete explanation of quantum theory. Yet, such approaches are only pursued outside of the mainstream, by those who do not fear the tarnishing of their academic credentials. Orthodoxy seems to have it that the quantum is here to stay: that it is absolutely fundamental. This is striking in contrast to GR and the Standard Model, which are considered not to be fundamental.

\section{Non-fundamentality of GR}\label{GR}

The main motivation for not considering GR as fundamental is the presence of spacetime singularities. Singularity theorems of Penrose and Hawking \citep{Penrose1965, Hawking1973} show that singularities are unavoidable in the theory under very general, physical reasonable conditions. There is a worry that these singularities represent the ``breakdown of spacetime''---or, at the very least---a breakdown of the theory. The latter interpretation is the standard one: the thought is that GR leaves something unexplained---what happens in those ``domains'' where the theory ``breaks down'' and ceases to be predictive? Thus, we are motivated to find a new, more-fundamental theory that does explain this.

Yet, it is not always clear exactly if, how, or why, spacetime singularities in GR signal a ``breakdown'' or incompleteness of the theory---and thus, whether, or how, they in fact serve as motivations for a more-fundamental theory \citep[][\S2.2]{SEPsing}. One way in which singularities have been argued as indicating the incompleteness of GR is through arguments that they lead to indeterminism \citep{Azhar2025, Earman1995}. Does this indeterminism require a new, more fundamental theory for its resolution? This depends, of course, on a number of factors, such as (i) how bothered we are by the indeterminism (if it affects our application or conceptualisation of the theory, if the indeterminism is a `physical' problem, \citep{Azhar2025, Earman1995}), (ii) whether it occurs in a regime where we'd expect more-fundamental quantum-gravitational effects to play a role (and thus that the new, deeper theory would be able to cure the problems for us), and (iii) whether there are other possibilities for resolving these pathologies. 

Regarding (ii), \citet{Weatherall2023} investigates the regimes ``where'' these pathologies occur, in order to figure out if they correspond to the ``high-energy'' regimes where we expect quantum gravity to take over. (That quantum gravity is associated with such regimes, though, is another assumption to question; discussed below, \S\ref{QG}). \citet{Weatherall2023} finds that, although not all such pathologies occur in this regime, there are some that do (and that these are the most likely to be physically salient), and thus could be expected to be resolved by quantum gravity. Regarding (iii) \citet{CrowtherDeHaro} show that there are some options for spacetime singularities other than resolving them by a more fundamental theory. The point is that singularities, even if they lead to an indeterminism in the theory, may not automatically require resolution, or resolution by quantum gravity. Yet, most physicists and philosophers are of the opinion that they are to be resolved by quantum gravity. I've spent some time discussing indeterminism here because, of course, there is also indeterminism associated with quantum theory---and, unlike in the case of GR---this indeterminism is not thought to be resolved by a more-fundamental theory. This is curious if we take quantum gravity to be a quantum theory, since it would seem we are trading indeterminism in one theory for indeterminism in another---where, if it is viewed as a problem, is to be resolved by interpretation or alternative approaches. 

There are also the empirical problems of dark matter and dark energy, which may signal  problems with GR, and the need for new theory to explain them---yet, standardly, they are not treated as such. It is interesting to ask why these are not usually considered problems motivating the need for a more-fundamental theory, since the measurement problem in quantum theory is also an empirical problem (or in any case, is not a purely theoretical or conceptual problem) not taken to motivate the need for a more-fundamental theory. The reason for this, in both cases, I think, is that the problems arise in ``macro'' regimes, and this is not where we typically expect a more-fundamental theory---especially not quantum gravity---to be necessary or applicable. Again, though, this attitude seems to rest on the mistaken idea of a more-fundamental theory being a ``micro'' theory, and the idea that quantum gravity only makes itself known at the Planck scale. Opening our minds a bit here may help us make progress on all these problems.

Other motivations for not treating GR as fundamental have to do with its not taking into account the quantum nature of matter. Additionally, there is the desire for a more unified theory. Again, these motivations prioritise the fundamentality of quantum theory, since it is not thought to be likewise modified by gravity, nor replaced by a more-fundamental theory.

\section{Non-fundamentality of QFT}\label{QFT}

Like quantum mechanics, QFT is a framework theory, within which many specific theories can be formulated. Popularly, it's viewed as a relativistic extension of quantum mechanics (though see below for considerations against this view), capable of handling systems with many `particles'. Standard quantum mechanics is unable to deal with large numbers of degrees of freedom, which is required if we are to describe spatially-extended bodies such as fluids (in condensed matter physics) and fields (such as the quantised electromagnetic field, which require a description of infinite degrees of freedom); it is also unable to deal with variable particle number (which is implied by the combination of quantum mechanics and special relativity, and necessary for describing scattering processes, where particles of different types can be created and destroyed). QFT is needed, also, if we are to have a more unified theory which takes into account the requirement, from relativity theory, that the laws of nature be Lorentz invariant (this requirement being in conflict with the Schr\"{o}dinger equation describing a system's dynamics in quantum mechanics). Thus motivates the new framework of QFT \citep{QFTSEP, Tong2006}.

Owing to its greater unification and broader domain of applicability, we can view QFT as a more-fundamental framework than quantum mechanics. Yet, it is based on familiar techniques from quantum theory, especially \textit{quantisation}. For this reason, a common view is that QFT is not a separate, more-fundamental framework than quantum mechanics, but rather a branch of quantum mechanics, or an extension of it (which is able to describe infinitely many degrees of freedom). What makes something a \textit{quantum} theory, and how much the framework of quantum mechanics needs to be modified before it is a new framework is an interesting question, which is also relevant when we are discussing quantum gravity (and the question of whether this need be a quantum theory, or whether it involves modifying quantum mechanics). Given my definition of a ``more fundamental theory'', above, I find it more useful to think of QFT as a new framework rather than a branch of quantum mechanics---some considerations which could be used in arguing for this are found in \citet[][\S1]{QFTSEP}, but I won't discuss these here.

We must distinguish between the framework of QFT (QFT---here I am discussing `conventional QFT')\footnote{\citet{Wallace2011}.}, and particular QFTs, such as the Standard Model. Neither of these are considered fundamental, even if they are more fundamental than standard quantum mechanics. There are several reasons why the framework of QFT is not considered fundamental. Primary among these is the much-discussed feature that it is mathematically ill-defined \citep{Fraser2020}. This leads to the second problem, which is that theories formulated in this perturbative framework were, traditionally, `plagued with ultraviolet divergences' \citep{Cao1993}. Historically, these infinities were removed in particular theories, such as quantum electrodynamics, via the procedure of renormalisation which rendered the theory finite and impressively predictive. This procedure, however, was physically suspicious, and the perturbative approach to QFT itself remained conceptually problematic. This changed with the development of renormalisation group techniques, which led to the \textit{effective field theory} framework for studying QFT systems at different energy scales, and ultimately to the discovery of the Standard Model. Now, the dominant philosophical interpretation of this QFT picture is that the ultraviolet divergences are not a real physical problem, but rather indications of the limitations of the framework of QFT. The framework itself is taken to be \textit{inherently approximate}, and its models are supposed to be \textit{effective theories}: not to be valid to arbitrarily high energy scales.\footnote{Cf. \cite{Fraser2020, Wallace2022, Wallace2011}.} 

There are also \textit{external reasons} for treating QFT as an ``effective framework'', rather than a fundamental one. The first is that although it takes into account special relativity, it is not \textit{generally relativistic}. Also worth mentioning is a problem to do with the derivation of various \textit{entropy bounds} in the context of black hole thermodynamics, which is a field of research which stems from attempts to bring together GR and QFT. Here, various theoretical results suggest that there is an upper bound on how much entropy can be contained in a given region of spacetime \citep{Bousso2002}. This striking result conflicts with the predictions of QFT (and indeed, any field theory), according to which the number of degrees of freedom in any given region of spacetime is infinite. Bousso takes this as strong reason not to expect that quantum gravity will be a QFT. 

The entropy bound---as with the related \textit{holographic principle}---plays a significant role in the search for a more fundamental theory of quantum gravity \citep[][\S7]{Crowtherbook}. Because the bound relates aspects of spacetime geometry to the number of quantum states of matter, the suggestion is that any theory that explains the bound will be a theory of quantum gravity: a theory that `combines' GR and QFT. The new theory of quantum gravity is expected to explain, via a reductive explanation, why it is that field theory (QFT) gives the wrong answer as to the amount of entropy able to be contained in a given region of spacetime. Interestingly, these results not only suggest the inapplicability of QFT at short distances (the expected domain of quantum gravity), but also at cosmologically large scales, calling into question the conventional analysis of the cosmological constant problem \citep[][p. 829]{Bousso2002}. (The cosmological constant problem is another problem which concerns both gravity and QFT, and is traditionally interpreted as indicating a failure of our renormalisation schemes to produce a finite vacuum energy consistent with the observational data). I mention this since it is a example of the utility of conceiving of quantum gravity as a more-fundamental theory (than QFT) according to my definition above: a theory with greater universality and broader domain of applicability, rather than (just) a more-micro theory of shorter length scales.

The two \textit{internal problems} mentioned at the start of this section, of the QFT framework being mathematically ill-defined and producing theories which require renormalisation, though, have also prompted an alternative response, \textit{axiomatic QFT} (AQFT), which is to attempt to set QFT on a firm, non-perturbative footing. Importantly, this approach is not an attempt at quantum gravity, or physics beyond QFT, but simply a new formulation of QFT at the level of QFT---i.e., as a combination of quantum mechanics and special relativity without any singularities in the theory.\footnote{See, e.g., \cite{Fraser2011, Fraser2009}.} The fact that AQFT is still an active field of research demonstrates that the first two internal problems mentioned above do not automatically motivate treating QFT as non-fundamental. And, if AQFT were successful, it would demonstrate that the interpretation of the divergences as indicative of our ignorance of new, more fundamental physics beyond, is mistaken. However, the \textit{external motivations} for treating this framework as non-fundamental---having to do with its relationship to GR---would nevertheless remain. It is this fact, as well as the other motivations for quantum gravity, that lead many philosophers/physicists to dismiss other possible solutions to the internal problems with QFT (or simply to ignore the problems,---as is the standard response, to e.g., the existence of Landau Poles in the Standard Model: they occur at a scale where we expect a new theory to take over anyway). 

I mention this because the internal problems in QFT---as with the spacetime singularities in GR---can motivate either a more-fundamental theory, or a solution ``at the same level''. But in both of these cases, the favoured approach is to move to a more fundamental theory. This is in contrast with the measurement problem in quantum mechanics: this too can motivate a solution either by a more fundamental theory, or else a new approach or interpretation ``at the same level'' as quantum theory. Yet, in this case, the favoured solution is the latter rather than the former.

Next, consider the Standard Model. This is the theory of all known elementary particles and forces; they are described as excitations of quantum fields in the framework of QFT (and it does not include gravity). The first reason for considering this theory non-fundamental is that it is a product of a non-fundamental framework. It is also non-unified: although the Standard Model can be written as a single theory, it appears as a disjointed amalgam of separate (particle) fields, which drives many physicists to seek a more unified theory beyond. These problems, as well as the problems of dark matter and dark energy, and the problem of quantum gravity, may motivate the need for a more fundamental physical theory. Other concerns with the Standard Model which drive physicists to search for `beyond Standard Model' physics, though, such as issues to do with naturalness, neutrino mass, matter-antimatter asymmetry, and so on, may also motivate a more fundamental theory \citep[See,][for a summary of these problems]{beyondSM}. In spite if this, these problems typically drive the search for a newer QFT valid at higher energy scales---evidence of which is to be found by building ever more-powerful particle colliders---rather than a new framework. This is not to say that a newer, more unified QFT with a broader domain of applicability would not be more fundamental than the Standard Model, but just that a theory that does all this and also \textit{reductively explains} the success of the framework of QFT would be more fundamental, and more explanatory still.

\section{Quantum gravity}\label{QG}

It is important to emphasise, at the outset, that we do not have a theory of quantum gravity. Though the need for this theory was recognised soon after the development of GR, physicists are still searching for quantum gravity (meaning it is a problem which has persisted longer than the 100 years of quantum interpretations). Here, I take `quantum gravity' to refer to any theory that satisfies the \textit{Primary Motivation} for quantum gravity, basically: a theory that describes the domains where both GR and QFT are believed to be \textit{necessary} (or, rather, where both theories are believed to be \textit{inaccurate}), and which somehow `takes into account' the lessons of both GR and quantum theory \citep{Crowtherbook}. Usually, the domains of quantum gravity are characterised as high-energy: where the perturbative QFT of gravity breaks down, at the Planck scale, as well as the interior of black holes and very early universe cosmology. While a theory of quantum gravity must (here, by definition) describe these domains, one of the points I'd like to emphasise in this chapter is that recognising that quantum gravity might also have large-scale consequences can be helpful for `broadening our horizons' in the search for the theory. This is emphasised by my new definition of relative fundamentality, according to which it is not by virtue of describing high-energy, or giving a micro-description of spacetime that quantum gravity is to be understood as a more-fundamental theory than GR and QFT (though, of course, it is expected to do this)---it is by being a more unificatory theory, with a broader domain of applicability than either of those theories. 

This leads to the second point, which is that quantum gravity somehow `combines' these GR and QFT. One way of doing this is via what \citet[][p. 33]{Crowtherbook} calls the \textit{New Framework Perspective}. This recognises both GR and quantum mechanics as non-fundamental and seeks (1) a new body of rules (the new theoretical framework) that express some insights of both GR and quantum mechanics, and (2) to recover \textit{both} GR and QM as approximations in their respective domains (i.e., to provide a reductive explanation of both theories). The \textit{Standard Perspective}, however, is to treat quantum mechanics as fundamental and GR as not. This perspective seeks (1) a \textit{quantum theory}, and (2) from this, the recovery of GR as an approximation in the domains where we know GR is successful (on such a view, the recovery of quantum mechanics is not required). The dominance of this view, and its motivations, is something which this chapter aims to question.
 
It is possible that QG take into account quantum mechanics without itself being a (purely) quantum theory, or, in particular, being a theory in which GR is quantised. The most familiar way of thinking about this is along the lines of semiclassical gravity or other `hybrid' approaches, which attempt to couple the classical and the quantum as a sort of `amalgamation' as opposed to a proper unification (or `mongrel gravity' theory \citep{Mattingly2009}). Another example is\textit{entropic gravity} \citep{Jacobson1995, Verlinde2011}. But there are other options which also involve modifying both GR and quantum theory. A proper unification---which is what is apparently most coveted in physics\footnote{\citet{Maudlin1996}}---of GR and the quantum would also involve modifying both theories, rather than `demoting' GR in favour of the quantum. The New Framework Perspective countenances the possibility that the framework of quantum mechanics---as it stands---is not applicable in quantum gravity, and gets modified somehow.

\section{The measurement problem and quantum interpretations}\label{measur}

In quantum mechanics, a wavefunction is assigned to an isolated physical system and its dynamics is governed by the Schr\"{o}dinger equation, which describes linear evolution. If we assume that the wavefunction is a complete description of the physical system, the linear dynamics is apparently incompatible with the appearance of definite results of measurements on the system. \citet{Maudlin1995} frames this form of the measurement problem as an incompatibility between the three claims:

\begin{itemize}
    \item (C1) the wavefunction of a physical system is a complete description of the system;
    \item (C2) the wavefunction always evolves in accord with a linear dynamical equation, e.g. the Schrödinger equation, which means that superpositions of possible solutions are also solutions;
\item (C3) each measurement has a definite result (which is one of the possible measurement results whose probability distribution satisfies the Born rule), or, in other words, superpositions of macroscopic objects are never observed.

\end{itemize}

In this way, \citet{Maudlin1995} provides a clear categorisation of the three general representationalist (`realist') ways of solving the measurement problem: hidden-variables approaches, dynamical collapse approaches, and the many-worlds interpretation. All of these seek to \textit{explain} why it is that we have this apparent inconsistency in quantum theory, and to do so by providing a story that gives us an understanding of what the world must actually be like given the success of this theory. Standardly, most of these attempt to solve the measurement problem at the level of quantum theory, and they do not offer us a more fundamental explanation of quantum theory. This is striking, as I've said, given that problems with GR and QFT do tend to motivate a deeper theory, and that we have reasons for seeking a deeper, more unified, theory. \textit{Nevertheless, some of these approaches and interpretations offer more heuristic and unificatory potential than others, which means they could help point the way to quantum gravity. I argue that this is a reason to favour such approaches, since it affords additional utility than merely offering us a picture of the world described by quantum mechanics.} The value is twofold: it may be that quantum gravity is a quantum theory, in which case your favoured approach should be applicable. But it may be that quantum gravity is a radically new theory, which your approach could help us find, provided it has good heuristic and unificatory potential. We have more to gain the less dogmatic we are, and the less committed we are to taking quantum theory as given fundamentally. 

\subsection{Bohmian mechanics}\label{Bohm}

The first approach involves denying (C1), and postulating that there are in fact additional variables and dynamics underlying the standard quantum description which are not reflected in the wavefunction. These are called \textit{hidden variables approaches} or \textit{additional variables approaches}, the best known being Bohmian mechanics, also known as pilot-wave theory \citep{GoldsteinSEP}. Proponents of this approach seek a \textit{constructive} understanding of quantum theory, and emphasise an ontology of particles which move in familiar 3-dimensional space \citep{Allori2023}. The particles are characterised by their positions, which evolve according to the `guiding equation', which expresses the velocities of the particles in terms of the wavefunction. Thus, the theory modifies standard quantum theory by the addition of a second equation of motion supplementing the Schr\"{o}dinger equation. Bohmian mechanics resolves the indeterminism in quantum theory, since the evolution is deterministic. Because it includes an additional dynamical equation and expands the state space of standard quantum mechanics, Bohmian mechanics is an alternative approach to quantum mechanics (a new theory) rather than an interpretation of quantum mechanics. The theory reproduces the results of standard quantum mechanics, and does not make any new testable predictions.

If this approach were accepted as a successful replacement for quantum theory, it would provide horizontally-reductive explanation of quantum theory, in terms of a newer theory which is more accurate. It would also provide a micro-constitutive explanation, providing a clear understanding of `what's going on', since it allows us to picture the micro physics in a similar way as we do the macroscopic realm of objects moving in 3-dimensional space, or 4-dimensional spacetime \citep{Allori2023, Allori2013}. But, as I have explained (\S\ref{sect:fund}), this would not be a deeper explanation of quantum theory by a more fundamental theory. This is also true of the many---but not all---of the dynamical-collapse approaches.

So far, most of the literature on both the Bohmian and dynamical-collapse approaches works with the strategy of modifying or supplementing the specific quantum theory of non-relativistic particle mechanics, and it is difficult to formulate QFT-versions of either Bohmian mechanics or dynamical-collapse models \citep[][p. 93]{Wallace2020}. \citet[][p. 94]{Wallace2020} criticises these approaches for their strategy of focusing on explaining only non-relativistic quantum particle mechanics, apparently done under the fiction of this being a fundamental universal theory. He interprets this charitably as being a bit of a warm-up exercise, `` `nonrelativistic quantum particle mechanics' is [taken as] a standin for the real fundamental quantum theory, and we [the Bohmians and dynamical-collapse theorists] hope that as many as possible of our conclusions carry over to hoped-for hidden-variables or dynamical-collapse theories formulated for the real fundamental theory''. The main argument in that paper, however, is even stronger: that these approaches---unlike the Everettian interpretation---are unable to account for the framework of quantum theory, which involves considering the relationships between the multiplicity of specific quantum theories used by physicists. Only the Everettian interpretation, Wallace argues, provides a self-consistent interpretive recipe for doing so. I return to this criticism below, arguing that it can provide reason to favour these approaches over the Everettian interpretation.

For now, notice that both the Bohmian and the dynamical-collapse approaches have aspirations of developing relativistic accounts. There are `relativistic' Bohmian theories as well as `relativistic' dynamical-collapse theories which have Lorentz invariant laws, but the former introduce a non-relativistic spatiotemporal structure (a preferred foliation), while the latter do not. For this reason, the dynamical-collapse approaches are usually claimed to be more compatible with relativity than the Bohmian approaches \citep[but cf.][against this]{Allori2022}. 

Bohmians, staunchly committed to a micro-constitutive ontology for quantum theory, are quite comfortable with thinking of relativity as non-fundamental: the preferred foliation is a necessary consequence of a `relativistic' Bohmian theory. \citet{Andrea2025} argues that both standard quantum theory and GR are incomplete theories: the former because it lacks a spatiotemporal ontology, and the latter because it does not accommodate the preferred foliation required for providing the spatiotemporal ontology of quantum theory (being, of course, the micro-constitutive ontology of Bohmian mechanics). Bohmians favour the fundamentality of their theory over both standard quantum theory and relativity on account of their theory being a \textit{constructive theory} of micro-ontology, compared to standard quantum theory and relativity, which are principle theories.

This modification of both quantum theory and relativity is an interesting proposal, but it doesn't yet reach the more-fundamental level of quantum gravity. \citet[][p. 79]{Allori2022} says that ``adding a foliation, which amounts to add an absolute frame, is an actual cost only if you think of relativity as fundamental. If one is willing to accept instead that relativity is not fundamental, then these [pilot wave] theories are very natural, as a preferred foliation is already in the cards due to nonlocality.'' The idea seems to be that, by taking Bohmian mechanics as more fundamental than GR, we have a modification of GR at the level of quantum theory (Bohmian mechanics). This gives us neither a more fundamental explanation of quantum theory, nor of gravity. We also still have classical spacetime on this proposal, which is significant for adherents of \textit{primitive ontology}: particles in 3-dimensional space or spacetime.

But, there are proposals for developing Bohmian versions of quantum gravity, which signify the heuristic potential of this approach---particularly because it opens up more possibilities for quantum gravity, including retaining continuum spacetime, as in \citet{Goldstein2001}. Or, developing a Bohmian version of existing attempts at quantum gravity, including loop quantum gravity \citep{BohmLQG}. The heuristic potential of the Bohmian approach for a more-fundamental theory seems to come from what \citet[][p. 97]{Wallace2020} sees as a defect of this approach in regards to its possibility of explaining quantum theory: that it fails to provide a consistent recipe of making sense of the multiplicity of quantum theories currently used by physicists. I suggest that this can be seen as a benefit of the approach: it is in an exploratory phase as regards quantum gravity---as it should be! Also, if the approach were to lead us to a theory of quantum gravity from which the less-fundamental quantum theories could be recovered, then this could potentially provide a deeper explanation of our current quantum theories.

\subsection{Dynamical Collapse Approaches}\label{GRW}

The \textit{dynamical-collapse theories} solve the measurement problem by denying claim (C2), that the wavefunction always evolves linearly; they revise the Schr\"{o}dinger equation by adding some nonlinear and stochastic evolution terms to explain the appearance of definite measurement results. The best known such approach is the GRW theory, following Ghirardi, Rimini and Weber \citeyearpar{GRW}. In this theory, the collapse is `built in' to the dynamical law: the wavefunction evolves according to the Schr\"{o}dinger equation up to a random time, at which the wavefunction centres around a random location. Since the rate of collapse is proportional to the number of `particles' in the system, macroscopic objects localise almost immediately. Unlike the Bohmian theory, the GRW theory does make differing predictions than standard quantum theory. Additionally, it is indeterministic, and does not postulate fundamental particles---rather `particles' are aspects of the behaviour of the universal wavefunction. There are debates over the ontology of the theory, which yield different versions, or interpretations, of it. All of these aim to give us a `picture' of what the theory tells us the world is like, and thus to provide some `representationalist' understanding of quantum theory in addition to solving the measurement problem. Again, though, they do not standardly give us a deeper, more fundamental theory.

These approaches, do, however, offer us excellent heuristic potential to move towards such a theory, by allowing us to explore the possibility that gravity plays a role in the collapse of the wavefunction, as suggested by \citet{Diosi, Karol}. In these approaches, it is not the number of particles which is significant for when the deviations from linear evolution become apparent, but rather the mass, or mass distribution of the system. Notably, Penrose adopts this view in his attempt to find a more fundamental theory of quantum gravity---one in which quantum mechanics is not taken as fundamental. Penrose adopts what I've called the New Framework Perspective, envisioning quantum gravity as being a radical departure from quantum theory. He believes that GR supplies clues as to the required modifications of quantum theory: we should not look to the standard perspective of quantising gravity, but rather to ``gravitise the quantum'', which means at least finding a non-linear theory to replace quantum mechanics \citep[][p. 817]{Penrose2004}. Such a theory---if we had it---would have the potential to provide a deeper explanation of the success, form, and features of both GR and quantum theory. Thus, the dynamical-collapse approaches have excellent heuristic value, and unificatory potential, in allowing us to explore the interaction between gravity and the quantum---particularly because they allow us to do so without necessarily going down to the Planck scale or the traditional domains of quantum gravity.

\subsection{Many-Worlds Interpretation}\label{Everett}

Finally, the third main representationalist proposal for solving the measurement problem involves denying the claim (C3), that each measurement has a single definite result. The \textit{Everettian} or \textit{many-worlds interpretation} says that there exists a multitude of equally real worlds, so that all possible results of measurements are realised \citep{Everett1957, DeWitt1973}. This interpretation is promoted as being the most mathematically straightforward way of understanding quantum theory, not modifying or supplementing the Schr\"{o}dinger equation, while being both local and deterministic. Everettians aim at reading physics at face value, thereby conceiving of quantum theory as a framework, able to encompass the many specific quantum theories that are formulated and utilised by physicists \citep{Wallace2020}. It explains the success of quantum theory in all its forms, by giving us a picture of what the world is like according to quantum theory---though this picture is rather extravagant and difficult to actually ``picture''. 

Do these features mean that the Everettian interpretation offers us a more fundamental explanation of quantum theory? It explains more than standard quantum theory, in that it explains also what happens in other worlds---but in the absence of independent reasons to suspect that these worlds exist, we would otherwise not require an account of their physics unless we subscribed to this interpretation.\footnote{This is similar to the case with the Bohmian ``actual positions of particles''.} It gives us an explanation of the framework of quantum theory, rather than just a few of its specific instances, though, so it \textit{does} offer a more fundamental explanation of the specific quantum theories used by physicists, in terms of their being part of the broader framework of quantum theory. But, the main selling point of this interpretation is that it doesn't say anything apart from what quantum theory says: it takes the theory at face value and attempts to interpret it literally \citep[][p. 13, 45]{Wallace2012}. Thus, the Everettian interpretation does not offer an explanation of quantum theory in terms of a deeper, more fundamental theory. The explanation is at the same level of fundamentality as standard quantum theory, conceived of as a framework theory.

Because the Everettian interpretation applies to the framework of quantum theory and provides a recipe for making sense of specific quantum theories, including specific QFTs, it is able also to be applied to quantum gravity provided that quantum gravity is a quantum theory in the usual sense. It takes the framework of quantum theory as fundamental---it is not going to be modified by any more fundamental theory---and thus the Everettian interpretation is also not supposed to be replaced by any more fundamental theory. While this puts the Everettians in a stable position, it places strong restrictions on the form of quantum gravity. Thus, this interpretation, which promotes itself for greater applicability and utility than the other approaches mentioned above, is apparently of rather limited potential if we are interested in a deeper, more fundamental explanation of quantum theory and relativity. I suggest that this is a reason to favour other approaches which are of \textit{more heuristic value} in the search for a new theory, and which \textit{open up more possibilities} as to the form of quantum gravity. 

The Everettian interpretation does offer some heuristic potential, however---an example is \citet{Boussomulti}, which argues that the many worlds of quantum mechanics and the many worlds of the multiverse of eternal inflation in cosmology are the same. This approach is a modification of the typical Everettian picture, but one which is promoted as being able to explain some otherwise puzzling features of our universe, such as the cosmological constant and dark energy. It is also unificatory in that it identifies the many worlds of quantum mechanics with the cosmological multiverse, so we no longer have two different scenarios offering us of infinite numbers of worlds, but just the one. (But, of course, its questionable whether we need either of these ontologies to begin with). Another example may be the \textit{gravitationally induced entanglement} experiments, which have been framed assuming an Everettian interpretation \citep{huggettQGLab}.

I've suggested that the many-worlds interpretation---while it may be applicable to quantum gravity if quantum gravity is not a theory which modifies quantum mechanics---limits us in several respects: 1) it is of limited heuristic potential for helping us find the new theory of quantum gravity; 2) it limits the form of what quantum gravity will look like; 3) it offers a shallower explanation of quantum theory than approaches which seek to explain quantum theory by a more fundamental theory. These aspects are due to its taking quantum mechanics at face value, which seems to prevent us from seeing potential avenues for unification of gravity and the quantum. This is really quite striking, of course, because \textit{the metaphysics proposed by many-worlds should have an impact on gravity}, or, rather, gravity (or whatever its quantum ``micro-ontology'') should have an impact upon the many worlds. So, the approach ought to offer unificatory potential, and excellent potential for yielding deep, novel insights into the fundamental nature of the universe: the interaction between gravity and the quantum. Yet, as far as I can tell, this has not been much pursued---much is being said about, e.g., energy conservation, locality and causality, whether there is local or global branching of worlds in connection with relativistic constraints, etc.,---but as to what the role of gravity is, or what the effect on gravity is, I'm having difficulty finding any literature (apart from that pertaining to the global multiverse, and gravitationally induced entanglement experiments mentioned above).\footnote{Cf. \citet[][pp. 311-312]{Wallace2012}.}

\section{Going deeper than quantum theory}\label{deeper}

A more fundamental theory than the Standard Model would be one which incorporates gravity, or other new `particles'. But, going deeper than QFT itself means moving to a more unified theory with a broader domain of applicability than the framework of QFT (note: `more unified' need not mean the theory is actually unified, but at least that it takes into account lessons from other areas of physics). Quantum gravity is capable of doing this, regardless of whether it proceeds via a quantisation of gravity, or whether it modifies the framework of quantum theory.\footnote{Note, that not all approaches to quantum gravity result in a more fundamental theory than QFT; \textit{asymptotic safety} is an example which, while being more fundamental than the Standard Model, would not produce a theory that is more fundamental than the framework of QFT.} Most approaches to quantum gravity involve the quantisation of GR (though with different choices of recipe). Following the path of problem-solving that results from this has led to some approaches now being quite different from standard QFT. String theory is an example of this, where we have a more fundamental framework than QFT, and one which reductively explains the framework of QFT as well as several of our particular QFTs. Yet, all these \textit{quantum} approaches to quantum gravity still face the measurement problem, and must look to interpretation for a solution.

The quantisation of gravity has been viewed as the natural means by which to unite gravity and the quantum---but, it signals the common view that the rules of QFT are ``immutable'' and it is GR which must ``bend itself appropriately to fit into the standard quantum mould'' (as \citet[][p. 817]{Penrose2004} puts it). Not having yet found a full, acceptable, theory of quantum gravity may suggest that a new strategy is useful, that perhaps quantisation is not the most fruitful starting point, and that more aspects of quantum (field) theory need to be modified. As Penrose says, GR may give us insights into which aspects these are. The representationalist approaches (rather than interpretations) to quantum mechanics also suggest that quantum theory requires supplementation or modification, but not for reasons of taking into account gravity or finding a more-fundamental theory---rather, for the reason of us making sense of it, which requires solving the measurement problem. Still, these approaches do not seek a more fundamental explanation of quantum theory---they do not go deeper. If we were to find a solution at the deeper level, though, this would give us a broader, more unified, and more-satisfactory explanation.

\citet[][pp. 215-216]{SmolinEinstein} sees this as the biggest reason to be sceptical of both pilot wave theory and collapse models, ``they make little contact with the other big questions in physics, such as quantum gravity and unification''. Viewing the problems in quantum foundations and the problem of quantum gravity as two sides of the same coin, Smolin---like Penrose---counsels us to look to GR to see how the quantum should be modified. The \textit{causal theory of views} is Smolin's proposal for a new framework which is more fundamental than, and explanatory of, both GR and QFT. Gravity is not quantised in this approach, though spacetime is emergent from discrete fundamental elements; it provides a solution to the measurement problem \citep[][p. 248]{SmolinEinstein} as well as a reductive explanation of non-relativistic quantum mechanics \citep[][p. 240]{Smolin2018, SmolinEinstein}.

Famously, Einstein did not believe that quantum theory was complete, and sought a solution to what he saw as its problems in a more fundamental theory which would unify electromagnetism with gravity (the two theories which described all the then-known forces of nature): a ``unified field theory'' from which the results of quantum mechanics could be derived, and thus be reductively explained. This was an unsuccessful endeavour in the later stages of his career, which isolated Einstein from the physics community as it forged ahead with the application and development of quantum theory. Now, we have our own great mavericks, in the later stages of their careers, seeking a solution in a more unified, more fundamental theory, while the rest of the community remains enchanted with the finality of the quantum. Could Penrose and Smolin be giving us bad advice here, and we're better to embrace interpretation because the quantum as we know it is here to stay? Well, of course it will stay so long as no-one dares follow their invitation to look deeper.\footnote{Paul Davies is another prominent physicist open to exploring the suggestion of modifying quantum theory, see \citet{Davies2025}.} So, the question becomes: how much do you look forward to another 100 years of quantum interpretation?

\section{Is there anything deeper than quantum theory?}\label{sceptic}

Why expect that the measurement problem is not to be solved by a more fundamental theory? Why is it to be `carried over' into quantum gravity? Why does it not motivate the non-fundamentality of quantum theory, in the way that the problems with GR and QFT do motivate our taking those theories as non-fundamental?

Some suggestions at possible answers include: 

\begin{enumerate}
    \item[A1.] The measurement problem is a problem that occurs in the wrong domain to be solved by a more fundamental theory;
    \item[A2.] Quantum gravity will be a quantum theory, and so inherit the problems of quantum theory, rather than solve them;
    \item[A3.] There is no way of reductively explaining quantum theory by a more fundamental theory (``to recover the quantum we need to include the quantum'').
\end{enumerate}

I can only briefly comment on these here. 

The first answer (A1) rests on a misconception regarding what is meant by a more fundamental theory, as being one which applies to higher energy scales, and/or describes micro-constituents. The measurement problem is a macro problem, thus, it may be thought that we should not be looking for its solution in a more fundamental theory. I've suggested that, rather, we should understand more fundamental theories as of broader domain, greater universality, and being more unificatory than the less-fundamental theories which they reductively explain. This allows us to recognise, in the case of quantum gravity, a greater range of empirical problems as potentially quantum-gravitational. Both GR and quantum theory are supposed to be universal theories, unrestricted in their domains of applicability (the fact that in practice, we only usually ever need to use one of the other is not supposed to be of consequence for these theories \textit{in principle} universality). As far as we know, the domains of these theories do not only intersect at the Planck scale, even if this where we expect both theories to fail.\footnote{Cf. \citet{Jacobs2025}.} Many quantum objects do have mass and exist in spacetime. Gravity could play a role in the quantum state reduction (collapse of the wavefunction), as suggested by Penrose, \citeauthor{Diosi} and \citeauthor{Karol}

Also, while we suppose that GR and quantum theory are universal, we know, too, that often it's only once we have a newer theory that we discover the actual domain of applicability of our original theory (think of Newtonian mechanics as a universal theory, until special relativity showed it was not). It might only be once we have quantum gravity that we realise that quantum theory is not actually universal. This leads into the second answer. 

The second answer (A2), that quantum gravity will be a quantum theory, of course depends on what we mean by a quantum theory. Perhaps its one which we construct via quantisation, and we believe that the quantisation of gravity is necessary, owing to the universality of quantum theory: quantum theory says that every dynamical field must be quantised, and GR tells us that spacetime is a dynamical field, therefore spacetime must be quantised. Or, maybe we have been persuaded of this by the various thought experiments proposed to demonstrate that a semiclassical approximation to quantum gravity, or any other `hybrid' approach which couples classical gravity to quantum fields, is somehow nonsensical or inconsistent. The basic idea usually being that the uncertainty relations in the quantised system spread to (`infect') the coupled non-quantised system. (In the case of semiclassical gravity, the thought is that the uncertainty in the position of a quantised gravitating object would lead to quantum uncertainty in the gravitational field, so the gravitational field itself should be quantised). None of these compel the quantisation of gravity. This has been pointed out, both in the physics and philosophy literature.\footnote{See references in \citet{Crowtherbook}, \S 5.} We have options for developing a theory of quantum gravity without quantising gravity. 

Maybe we see QFT as a more fundamental framework than quantum mechanics, and we see that moving to QFT hasn't solved the measurement problem, so we are sceptical that moving to an even more fundamental framework still might not get us there. In the case of moving from quantum mechanics to QFT, though, we retained quantisation as the method of theory-generation. To repeat: we have options for developing a theory of quantum gravity without quantising gravity. We don't have to carry over the problems of quantum theory to a more fundamental framework---in fact, we could solve them with the new framework. As a bonus, we might even avoid some problems associated with quantisation, such as the problem of time. 

Finally, (A3) has been suggested since it seems there are aspects of quantum theory which resist any deeper explanation. We accept that quantum theory tells us that the world is non-local (or that superpositions exist, or whatever other feature of the theory we see as providing essential insight into the world) at least at the level we know is accurately described by quantum theory (i.e., the domains we know that the theory works). Then, it might seem puzzling that a more fundamental theory could be local one (or one without superposition, etc.) How could we recover this feature of the theory, if it were not present in a more fundamental theory? A few comments on this. First, there will likely be features of quantum theory that are retained in quantum gravity, even if this theory is a New Framework. 

Also, we have various interpretations and approaches to quantum theory which argue that we can explain, e.g., indeterminism with a deterministic theory, or non-locality in terms of a local theory, etc. So, the claim that we cannot recover these features from a more fundamental, non-quantum theory requires a more specific and detailed argument. Maybe it's just weird to think that we could have locality at the classical level, but not at the more-fundamental quantum level, but then, somehow locality is back in the more-fundamental theory of quantum gravity beyond that. Yes, that would be weird, but so long as we have derivations connecting these theories (which we would, since it is necessary for establishing that one theory is more fundamental than another), together with some story about how these derivations capture the \textit{physically salient} aspects of the theories, we can have a reductive explanation of how and why it is that the world happens to be structured this way.

Third, one might think the same about spacetime: that it cannot be recovered from a more fundamental non-spatiotemporal theory. Yet, many approaches to quantum gravity rely on the idea that we can do just that.

\section{Conclusion}

By offering a new conception of relative fundamentality, I hope to have shown that we can regard GR and quantum theory as equally fundamental, and both as being less fundamental than quantum gravity. This may help us see that, just as problems with GR motivate resolution by quantum gravity, so too can problems with quantum theory. Recognising that quantum gravity is more fundamental not because it applies at higher energy scales, or gives a micro-constitutive explanation, but because it has a broader domain, greater universality and unification, allows us to see, as well, that even ``macro'' problems may be quantum-gravitational problems. And, importantly, that finding quantum gravity may involve not only modifying GR, but also quantum theory.

Rather than explaining quantum theory, and solving the measurement problem by looking to interpretations of, or alternative approaches to, quantum theory, we might explain quantum theory in terms of a more fundamental theory, and find a solution to the measurement problem at the level of this more fundamental theory. This would offer a deeper, more satisfying explanation of quantum theory.

I've suggested a new way of assessing quantum interpretations and approaches, in terms of their heuristic potential in helping us find this more fundamental theory. The many-worlds interpretation comes out behind the pilot wave approaches and dynamical collapse approaches in this assessment. This is because, as an interpretation of quantum theory, it favours the fundamentality of quantum theory, and thus limits the possibility of finding deeper explanation in a more-fundamental theory. Given we've spent 100 years on quantum interpretations and 100 years searching for quantum gravity, we may want to open up our explorations---Penrose and Smolin have shown us that a way to do this is by considering these problems together, in a more unified way, and they encourage us to do so whilst we are a little farther away from being 100 years old ourselves.

\section{Acknowledgements}
Thanks to Christian Airikka and J\o rn Mjelva for discussions, and to Lars-G\"{o}ran Johansson and Rasmus Jaksland for feedback.

\bibliography{bibli}

\end{document}